\documentstyle[preprint,aps,prb,epsfig]{revtex}

\begin{document}

\title{Zero-energy peak of the density of states and localization
properties of a one-dimensional Frenkel exciton: Off-diagonal disorder}

\author{G.\ G.\ Kozlov and V.\ A.\ Malyshev}
\address{All-Russian Research Center "Vavilov State Optical Institute",
Birzhevaya Liniya 12, 199034 Saint-Petersburg, Russia}

\author{F.\ Dom\'{\i}nguez-Adame}
\address{GISC, Departamento de F\'{\i}sica de Materiales, Universidad
Complutense, E-28040 Madrid, Spain}

\author{A.\ Rodr\'{\i}guez}
\address{GISC, Departamento de Matem\'{a}tica Aplicada y
Estad\'{\i}stica, Universidad Polit\'{e}cnica, E-28040 Madrid, Spain}

\date{\today}

\maketitle

\begin{abstract}

We study a one-dimensional Frenkel Hamiltonian with off-diagonal
disorder, focusing our attention on the physical nature of the
zero-energy peak of the density of states.  The character of excitonic
states (localized or delocalized) is also examined in the vicinity of
this peak by means of the inverse participation ratio.  It is shown that
the state being responsible for the peak is localized.  A detailed
comparison of the nearest-neighbor approach with the long-range
dipole-dipole coupling is performed.

\end{abstract}


\narrowtext

\tighten

\section{Introduction} \label{Intro}

Since the pioneering works of Anderson,\cite{Anderson58} and Mott and
Twose,\cite{Mott61} electronic and transport properties of randomly
disordered systems have been the subject of long-lasting interest both
from fundamental and applied viewpoints.\cite{Anderson58,Mott61,Lieb66,%
Alexander81,Haus87} One-dimensional ($1$D) systems are frequently
considered because they turn out to be simpler than those in three
dimensions.\cite{Lieb66} Originally, Mott and Twose~\cite{Mott61}
conjectured that all states are localized in $1$D systems, for any
degree of disorder.  Afterwards, a great deal of work has been devoted
to examine Mott-Twose conjecture (see, for instance, the
review~\onlinecite{Thouless74}).  However, it is well-known that
electron delocalization appears in 1D random systems with short-range
correlations.\cite{Dunlap90,Sanchez94}

Two decades ago, Theodorou and Cohen established that the density of
states (DOS) of a $1$D tight-binding Hamiltonian with nearest-neighbor
(NN) interactions and random off-diagonal elements presents a
singularity at the center of the band.\cite{Theodorou76} These authors
used an analytical approach based on previous results obtained by
Dyson~\cite{Dyson53} for disordered linear chains of harmonic
oscillators.  In Ref.~\onlinecite{Theodorou76}, it was also stated that
the corresponding state is delocalized as the localization length was
found to be infinite.  Adding some amount of diagonal disorder in the
presence of off-diagonal randomness makes all states to be localized.
\cite{Economou71} Remarkably, the first calculations on $1$D
tight-binding Hamiltonians with only diagonal disorder did not reveal
any singularity neither in the DOS nor in the localization
length.\cite{Bush72,Papatriantafillou73} Further, it was found both
numerically \cite{Czychol81} and analytically \cite{Kappus81,Derrida84}
a very weak anomaly (a peak but not a singularity) in both properties
mentioned above.

Recently, Fidder {\em et al\/}.  have found by numerical diagonalization
of the $1$D Frenkel Hamiltonian with off-diagonal disorder that,
notwithstanding the singularity of the DOS, the corresponding state is
localized if one includes the long-range (LR) interactions due to
dipolar coupling between different sites.\cite{Fidder91} This finding
seems to be in contradiction with the point of view raised in
Ref.~\onlinecite{Theodorou76} suggesting that the state corresponding to
the singularity of the DOS is delocalized.  In this paper, we examine in
detail the conclusions of Refs.~\onlinecite{Theodorou76}
and~\onlinecite{Fidder91}.  We address this issue by considering a $1$D
Frenkel Hamiltonian with off-diagonal disorder with NN interactions and
compare the results with those obtained when LR interactions are taken
into account.  The rest of the paper is organized as follows.  In
Sec.~\ref{?}, the $1$D Frenkel Hamiltonian with NN interactions is
analyzed.  We present arguments against those raised in
Ref.~\onlinecite{Theodorou76}, namely that the zero-energy state is
localized, even in the NN problem.  This conclusion, based on analytical
considerations, is then confirmed by direct diagonalization of the
Hamiltonian.  The detailed study of the $1$D Frenkel Hamiltonian with LR
interactions is presented in Sec.~\ref{FrHam}.  Section~\ref{Simul}
deals with the numerical simulations confirming the analytical results.
Using numerical diagonalization of a Frenkel Hamiltonian with LR
interactions, we calculate both the DOS and the inverse participation
ratio, to be defined below, and study new features of these magnitudes
with respect to the NN approach.  Section~\ref{Sum} concludes with some
comments regarding the results we have obtained.

\section{Is the zero-energy state delocalized?} \label{?}

In this Section we briefly review the arguments of
Ref.~\onlinecite{Theodorou76} leading to the conclusion that the state
at the center of the band is delocalized.  We present further arguments
suggesting the opposite point of view and, what is most important,
numerical calculations confirm our statement.  Let us consider a
tight-binding Hamiltonian with only NN interactions
\begin{equation}
H = \sum_n U_{n,n+1} (|n\rangle\langle n+1| + |n+1\rangle\langle n|),
\label{H_NN}
\end{equation}
where the NN interactions $\{U_{n,n+1}\}$ are assumed to be
$\delta$-correlated and similarly distributed stochastic variables.  The
state vector $|n\rangle$ represents an excitation at {\em site\/} $n$.
All site energies are set to zero since no diagonal disorder is
included. The eigenvalue problem of the NN model reads
\begin{equation}
U_{n,n+1}a_{n+1} + U_{n,n-1}a_{n-1} = Ea_n,
\label{Eigen}
\end{equation}
where the set $\{a_n\}$ represents the real eigenvector corresponding to
the eigenenergy $E$.  For zero energy Eq.~(\ref{Eigen}) gives the
recurrence relation $a_{n+1}=-(U_{n,n-1}/U_{n,n+1})a_{n-1}$.  Using this
relation one can find
\begin{equation}
a_{2n+1} = \Biggl(-{U_{2n,2n-1}\over U_{2n,2n+1}}\Biggr)
\Biggl(-{U_{2n-2,2n-3}\over U_{2n-2,2n-1}}\Biggr)\ldots
\Biggl(-{U_{2,1}\over U_{2,3}}\Biggr) a_1.
\label{a}
\end{equation}
The amplitudes at even positions vanish. The eigenvector (\ref{a})
represents the zero-energy state for a chain with odd number of sites.
Defining the localization length at the center of the band $L(E=0)$ by
the expression
\begin{equation}
{1\over L(E=0)} = - \lim_{n\rightarrow\infty} {1\over 2n}
\ln\Biggl|{a_{2n+1}\over a_1}\Biggr|,
\label{L}
\end{equation}
and applying the central-limit theorem, the authors of
Ref.~\onlinecite{Theodorou76} obtained $1/L(E=0)=0$.  From this result
they concluded that the state at center of the band was extended.

The definition of the localization length (\ref{L}) is based on an
unconditional assumption of the so-called exponential localization.
Indeed, in such a case one would have typically $a_{2n+1} \sim
\exp[-(2n+1)/L]$.  Certainly, the definition (\ref{L}) cannot discern
between a weaker than exponentially localized state (where the amplitude
$a_{2n+1}$ decreases or increases with $n$ slower than an exponential)
and an extended state (where $a_{2n+1} \sim 1/\sqrt N$ with $N
\rightarrow\infty$ being the number of sites in the chain).  In such a
case, the mean extension of the eigenfunctions or the inverse
participation ratio are the more adequate quantities for learning the
character of the state.

The results presented below show that the problem we are discussing just
belongs to those which cannot be adequately analyzed from the
assumptions leading to Eq.~(\ref{L}).  First, let us write the NN
interactions in the form $U_{n,n+1} = U_0 (1 + \xi_{n,n+1})$, where
$\xi_{n,n+1}$ is a Gaussian distributed stochastic variables with
variance $\xi_0^2 \ll 1$.  Then, it is easy to calculate the probability
distribution of $\xi_{2n+1} \equiv \ln |a_{2n+1}/a_1|$,
\begin{equation}
g(\xi_{2n+1}) = {1 \over \sqrt {4\pi n}\xi_0}
\exp \Biggl (- {\xi_{2n+1}^2 \over 4\xi_0^2 n}\Biggr ).
\label{g_xi}
\end{equation}
{}From this, the authors of Ref.~\onlinecite{Lifshits82} claimed that
typically $|a_{2n+1}/a_1| \sim \exp(\pm 2\xi_0\sqrt n)$.  If so, one
should conclude that the zero-energy state is localized rather than
extended, in contradiction with the statement of
Ref.~\onlinecite{Theodorou76}.

Further, using Eq.~(\ref{g_xi}) we can calculate the probability
distribution of $\chi_{2n+1} \equiv |a_{2n+1}/a_1|$,
\begin{equation}
f(\chi_{2n+1}) = {1 \over \sqrt {4\pi n}\xi_0 \chi_{2n+1}}
\exp \Biggl (- {\ln^2\chi_{2n+1} \over 4\xi_0^2 n}\Biggr ).
\label{f_chi}
\end{equation}
This function has a sharp peak at $\chi_{\mathrm{max}} = \exp(-2 \xi_0^2n)$ 
and a very broad tail for large $\chi_{2n+1}$ such that
$\langle\chi_{2n+1}\rangle = \int \chi f(\chi) d\chi = \exp(\xi_0^2 n)$.
Thus, it is rather difficult to make a definite conclusion from
Eq.~(\ref{f_chi}) concerning a typical dependence of $|a_{2n+1}/a_1|$ on
$n$.  Nevertheless, the fact that $f(0) = 0$ certainly indicates the
zero probability to obtain an extended state.  Below, we confirm this
observation by numerical simulations.

\section{Frenkel Hamiltonian} \label{FrHam}

We will be also interested in studying both the DOS and the degree of
localization of states of a 1D tight-binding Hamiltonian including all
(LR) interactions, beyond the NN interactions.  According to this, we
then introduce the complete Hamiltonian
\begin{equation}
{\cal H}=\sum_{\stackrel{m,n=1}{m\neq n}}^{N}U_{mn}|m\rangle\langle n|,
\label{Hn}
\end{equation}
in which summation is performed now over all pair of sites.  For
definiteness, it is assumed hereafter that excitations described by the
presented Hamiltonian correspond to Frenkel excitons.  Furthermore, the
$U_{mn}$ is assumed to be of dipole-dipole nature.  We restrict
ourselves to the case in which all transition dipole moments have the
same magnitude and direction.  Thus, one can take $U_{mn} =
-U/|\xi_m-\xi_n|^3$, where $-U$ ($U>0$) is the dipole-dipole coupling of
nearest-neighbors in the periodic lattice, i.e.  at $\xi_m-\xi_{m+1}=1$
(we chose here the negative sign of NN coupling as it takes place, for
example, in J-aggregates~\cite{Fidder91}), and $\xi_m=m+\delta_m$ with
$\delta_m$ being a stochastic variables assuming to be distributed
around the regular sequence according to the Gaussian law with variance
$\sigma^2$
\begin{equation}
{\cal P}(\delta_m)=\Biggl({1\over 2\pi\sigma^2}\Biggr)^{1/2}
\exp\Biggl(-{\delta_m^2\over 2\sigma^2}\Biggr).
\label{Gauss}
\end{equation}

\subsection{The exciton spectrum and the DOS in the absence of disorder}
\label{A}

Before any discussions of the effects resulted from localization, it is
useful to recall the peculiar features of the 1D-exciton spectrum and of
the DOS in the absence of disorder $(\delta_m=0)$.  Then the Hamiltonian
(\ref{Hn}) can be approximately diagonalized (with accuracy of the order
of $N^{-1}$) by introducing the excitonic basis~\cite{Malyshev95}
\begin{equation}
|k\rangle = \left( {2\over N+1}\right)^{1/2}
\sum_{n=1}^N\sin\left({\pi kn\over N+1}\right)|n\rangle.
\label{k}
\end{equation}
The state vector $|k\rangle$ represents an exciton in the $k$-th state.
Substituting (\ref{k}) into Eq.~(\ref{Hn}) one obtains~\cite{Malyshev95}
\begin{mathletters}
\label{one}
\begin{equation}
{\cal H}=\sum_{k=1}^N E_k |k\rangle\langle k|,
\label{Hk}
\end{equation}
\begin{equation}
E_k=-2U\sum_{n=1}^N{1\over n^3}\cos\left({\pi kn\over N+1}\right) +
{\cal O}(N^{-1}).
\label{Ek}
\end{equation}
\end{mathletters}
Equation~(\ref{Ek}) generalizes the corresponding expression of the
NN approximation [$n=1$ term in (\ref{Ek})] to the case of including all
(LR) interactions.  We are especially interested in the behavior of the
spectrum and of the DOS in the vicinity of extreme points, $K\equiv\pi
k/(N+1)=0$ and $K=\pi$, as well as at the center of the band, $k =
(N+1)/2$ ($N$ taken to be odd).  To do that, we exploit the following
equation~\cite{Gradshtein80}
\begin{equation}
\sum_{n=1}^\infty {\cos Kn\over n}=-\ln\Biggl(2\sin {K\over 2}\Biggr)
\label{Grad}
\end{equation}
and the fact that one can extend the sum in Eq.(\ref{Ek}) up to infinity
because $n^{-3}$ decreases with $n$ fast enough.  Then, by integrating
Eq.~(\ref{Grad}) twice with respect to $K$, the sum in Eq.~(\ref{Ek})
can be cast into the form~\cite{Malyshev95}
\begin{mathletters}
\label{two}
\begin{equation}
E_k=-2U\zeta (3)+U\Biggl({3\over 2}-\ln K\Biggr)K^2 , \quad K\ll 1,
\label{Bottom}
\end{equation}
\begin{equation}
E_k={3\over 2}U\zeta (3)-U\ln 2(K-\pi)^2,\quad K-\pi\ll 1,
\label{Top}
\end{equation}
\end{mathletters}
where $\zeta (3)=\sum_{n=1}^\infty n^{-3}\approx 1.202$.  The
corresponding formulae within the NN approximation are $E_k=-2U+UK^2$ if
$K\ll 1$ and $E_k=2U-U(K-\pi)^2$ if $K-\pi\ll 1$.  Thus, one can
conclude that LR interactions affect both the position of the bottom and
the top of the band as well as the DOS of $1$D excitons.  As we can see,
the bottom and top of the band change, respectively, from $-2U$ to
$-2U\zeta (3)\approx -2.404U$ and from $2U$ to $(3/2)U\zeta (3)\approx
1.803U$.  The DOS decreases approximately by the factor $\sqrt{\ln |E|}$
in the vicinity of the bottom of the band and, on the contrary, grows by
the factor $1/\ln 2$ close the top.

Finally, we would like to comment on the energy of the central exciton
band state, with $k = (N+ 1)/2$.  In the NN model, one finds that
$E_{(N+1)/2} = 0$, while with including all dipolar couplings, this
energy is shifted to
\begin{equation}
E_{(N+1)/2} = -2U\sum_{n=1}^N {1\over n^3}
\cos\left({\pi n\over 2}\right) \approx 0.225 U.
\label{Epeak}
\end{equation}
The DOS in the vicinity of the band center does not change noticeably
as compared to the NN model.

\subsection{Motion narrowing effect} \label{B}

In the presence of disorder, the Hamiltonian of the system can be
written as a sum of two parts: the unperturbed one (\ref{one}) and
a term produced by the fluctuations of $U_{mn}$
\begin{mathletters}
\label{three}
\begin{equation}
{\cal H}=\sum_{k=1}^N E_k |k\rangle\langle k|+\sum_{k,k^\prime=1}^N
\Delta_{kk^\prime}|k\rangle\langle k^\prime|,
\label{Hp}
\end{equation}
\begin{equation}
\Delta_{kk^\prime} = {2\over N+1}\sum_{m,n=1}^N\delta U_{mn}
\sin\left({\pi kn\over N+1}\right)
\sin\left({\pi k^\prime m\over N+1}\right),
\label{Delta}
\end{equation}
\end{mathletters}
where $\delta U_{mn}=U_{mn}-{\bar U}$ where ${\bar U}$ means averaging
over the probability distribution (\ref{Gauss}).  Here
$\Delta_{kk^\prime}$ have diagonal and off-diagonal parts.  The former
is responsible only for the inhomogeneous broadening of excitonic
levels, while the latter couples the excitonic modes and, therefore,
causes the localization effects.

The $\Delta_{kk^\prime}$ undergo fluctuations because $\delta U_{mn}$
fluctuate.  Assuming NN coupling and that $\delta U_{mn}$ fluctuations
are small in some sense (see below), we can find the
$\Delta_{kk^\prime}$ distribution in an analytical form.  This also
helps us to comment on the results of numerical simulations that we
discuss later in Sec.~\ref{Simul}.

In order to achieve the task, we use the definition
\begin{eqnarray}
{\cal P}(\Delta_{kk^\prime}) =
\left<\delta\left(\Delta_{kk^\prime} - {2\over N+1}
\sum_{n=1}^{N-1}\delta U_{n,n+1}\left[\sin\left({\pi kn\over N+1}\right)
\sin\left({\pi k^\prime (n+1)\over N+1}\right)\right.\right.\right.
\nonumber\\
\nonumber\\ \left.\left.\left.
+\sin\left({\pi k^\prime n\over N+1}\right)
\sin\left({\pi k(n+1)\over N+1}\right)\right]\right)\right>.
\label{Def}
\end{eqnarray}
Here, angle brackets indicate the average of the $\delta$-function over the
fluctuations of NN distances.  They obey a Gaussian distribution law like
(\ref{Gauss}) but replacing $\sigma^{2}$ by $\sigma_{NN}^{2} =
2\sigma^{2}$.  We omit the details of tedious but straightforward
trigonometric calculations and only quote the final results.

It can be easily shown that the sum in Eq.~(\ref{Def}) is exactly equal to
zero if $k+k^\prime = N+1$.  Particularly, this means that
$\Delta_{kk^\prime}=0$ for $k=k^\prime=(N+1)/2$ when $N$ is taken to be
odd, i.e., the first order correction to the central energy is exactly
equal to zero and does not fluctuate.  Fluctuations of the other
$\Delta_{kk^\prime}$ are distributed according to the Gaussian function
with variances of the diagonal and off-diagonal elements distribution,
$\sigma_d^2(k)$ and $\sigma_{nd}^2(k,k^\prime)$, given by
\begin{mathletters}
\label{four}
\begin{equation}
\sigma_d^2(k)={(6\sigma U)^2\over N+1}
\Biggl[2+\cos\left({2\pi k\over N+1}\right)\Biggl], \quad k\ne \frac{N+1}{2}.
\label{sigmad}
\end{equation}
\begin{equation}
\sigma_{nd}^2(k,k^\prime)=
{(6\sigma U)^2\over N+1}\Biggl[1+\cos\left({\pi k\over N+1}\right)
\cos\left({\pi k^\prime\over N+1}\right)\Biggl], \quad k+k^\prime \ne N+1.
\label{sigmand}
\end{equation}
\end{mathletters}
{}From Eqs.~(\ref{four}) one can conclude that, in the case of off-diagonal
disorder, the motion narrowing effect is also present as it takes place for
diagonal disorder,\cite{Knapp84} i.e., both magnitudes in (\ref{four})
scale as $(N+1)^{-1}$.  We should point out that, in contrast to diagonal
disorder, here the magnitudes $\sigma_d$ and $\sigma_{nd}$ are functions of
the state numbers.  Note also that $\sigma_d(k)$ goes through its minimum
value exactly at the center of the exciton band, i.e., at $k=(N+1)/2$ and
$k=N/2$ at $N$ taken odd and even, respectively.  In fact, we can also
assert this with respect to the value of $\sigma_{nd} (k,k^\prime)$ since
$k$ and $k^\prime$ cannot differ greatly provided the condition
$\sigma_{nd}(k,k^\prime)\ll U$.

To conclude this Section let us comment on the validity of the motion
narrowing effects.  Obviously, this is valid only when $\sigma_{nd} <
|E_k-E_{k+1}|$.  In this case, the excitonic states are not mixed by the
perturbation and remain to be extended over the whole chain.  They are
essentially mixed for the opposite sign of inequality, reducing their
localization lengths.  Then, the number of sites within the region of
localization $(N^{*})$ drives the motion narrowing effect rather than
the whole number in the chain $N$.  In Refs.~\onlinecite{Malyshev95}
and~\onlinecite{Malyshev91} a self-consistent rule for estimation of
$N^*$ is carried out.

\section{Numerical simulations and discussions}
\label{Simul}

Further, we will mainly focus our attention on the normalized density of
states $\rho (E)$ and on the degree of localization (inverse
participation ratio, IPR) for the states at energy $E$.  They are
defined respectively as follows~\cite{Fidder91}
\begin{mathletters}
\begin{equation}
\rho (E) = {1\over N}\Biggl\langle\sum_k \delta(E - E_k)\Biggr\rangle,
\label{rho}
\end{equation}
\begin{equation}
{\cal L}(E) = {1\over N\rho(E)}\Biggl\langle\sum_k\delta(E - E_k)
\Biggl(\sum_{n=1}^N a_{kn}^4\Biggr)\Biggr\rangle,
\label{IPR}
\end{equation}
\end{mathletters}
where the angular brackets indicate an average over an ensemble of
disordered linear chains and the $a_{kn}$ is the eigenvector of the
Hamiltonian (\ref{Hn}) corresponding to the eigenvalue $E_k$ with
$k=1,2,3,\ldots,N,$
\begin{equation}
\sum_{m=1}^N U_{nm}a_{km} = E_k a_{kn}.
\label{eigen}
\end{equation}
The IPR behaves like $1/N$ for delocalized states spreading uniformly
over the entire system on increasing $N$.  In particular, the IPR can be
exactly computed for the eigenstates of the periodic lattices given in
(\ref{k}).  In doing so we obtain the expected behavior for $N\to
\infty$.  On the contrary, localized states exhibit much higher values.
In the extreme case, when the exciton is localized at a single site, the
IPR becomes unity.  Therefore, the scaling analysis of the IPR as a
function of the system size provides valuable information about the
nature of the excitonic states.  We should mention that a complete
multifractal analysis, accomplished by studying the scaling of the other
moments of the probability distribution, is beyond the scope of this
work.

\begin{figure}
\centerline{\epsfig{file=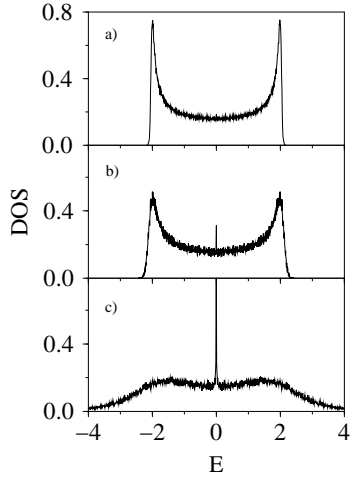,width=8cm}}
\caption{Density of states in the frame of NN coupling when the
lattice size is $N=2500$ and the degree of disorder is a)
$\sigma_{NN}=0.02$, b) $0.04$, and c) $0.16$.}
\label{fig1}
\end{figure}

We have solved numerically the eigenvalue problem~(\ref{eigen}) for
different values of disorder, namely the mean fluctuation of the NN
distance, $\sigma_{NN}=\sqrt{2}\sigma$, to study the features of both
the DOS and the IPR discussed above.  In our numerical treatment
$\sigma_{NN}$ ranges from $0$ (periodic lattices) up to $0.32$ whereas
the maximum system size we have considered is $N=2500$.  Results
comprise averages over $50$ realizations of the disorder for each given
pair of parameters $N$ and $\sigma_{NN}$.

\subsection{Nearest-neighbor approximation} \label{NN}

\begin{figure}
\centerline{\epsfig{file=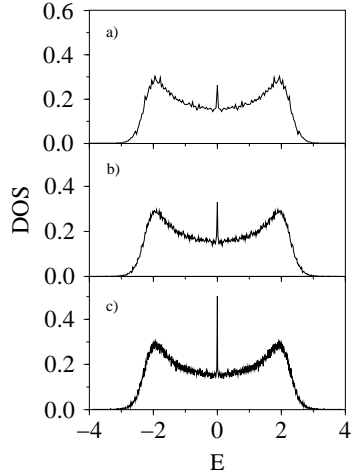,width=8cm}}
\caption{Density of states in the frame of NN coupling when the degree
of disorder is $\sigma_{NN}=0.08$ and the lattice lattice size is a)
$N=1000$, b) $1500$, and c) $2500$.}
\label{fig2}
\end{figure}

Let us comment the results we have obtained for the NN approximation.
Figure~\ref{fig1} shows the DOS for the largest lattice size we have
considered ($N=2500$) and different values of the disorder ($\sigma_{NN}
= 0.02,0.04,0.16$ from top to bottom).  We observe that the DOS is
symmetric about the center of the band.  The singularities at the edge
of the exciton band are smeared out on increasing the degree of
disorder.  Interestingly, a sharp peak in the DOS at the center of the
exciton band appears when the degree of disorder exceeds some threshold
value ($\sigma_{NN} \approx 0.03$ for our model parameters).  We will
discuss further this point later.  We have also observed that the
percentage of states in the DOS peak increases with the degree of
disorder.  In addition, the amplitude of the peak rises noticeably with
increasing the number of sites in the lattice, as seen in
Fig.~\ref{fig2} for $\sigma_{NN} = 0.08$.

\begin{figure}
\centerline{\epsfig{file=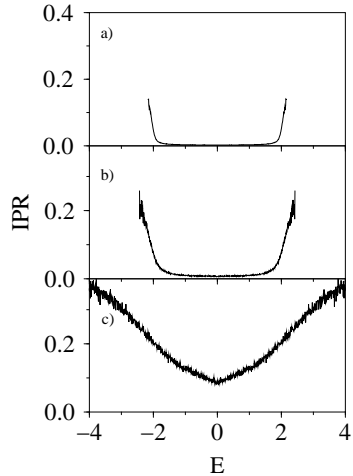,width=8cm}}
\caption{Inverse participation ratio for the same cases shown in
Fig.~\protect{\ref{fig1}}. Notice the overall increase on increasing
the degree of disorder.}
\label{fig3}
\end{figure}

\begin{figure}
\centerline{\epsfig{file=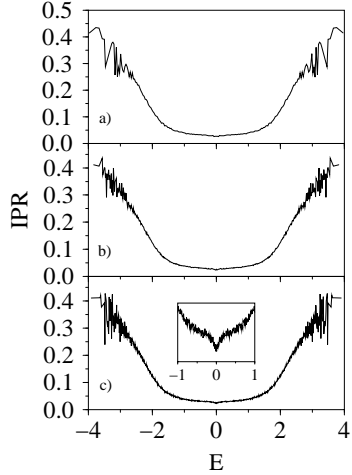,width=8cm}}
\caption{Inverse participation ratio for the same cases shown in
Fig.~\protect{\ref{fig2}}. The inset shows an enlarged view of the
center of the band.}
\label{fig4}
\end{figure}

The IPR presents an overall increase when the degree of disorder
increases, meaning that the larger the degree of disorder, the smaller
the exciton localization length.  This is clearly observed in
Fig.~\ref{fig3}, where we show the IPR as a function of the exciton
energy for the same parameters of Fig.~\ref{fig1}.  However, the
increase of the IPR strongly depends on the energy, being more
pronounced close to the center of the band.  Simultaneously with the
occurrence of the peak of the DOS, a hardly visible downfall arises in
the IPR at zero energy.  This downfall is better revealed for larger
lattices, as it can be seen in Fig.~\ref{fig4} for the same parameters
of Fig.~\ref{fig2}.

\begin{figure}
\centerline{\epsfig{file=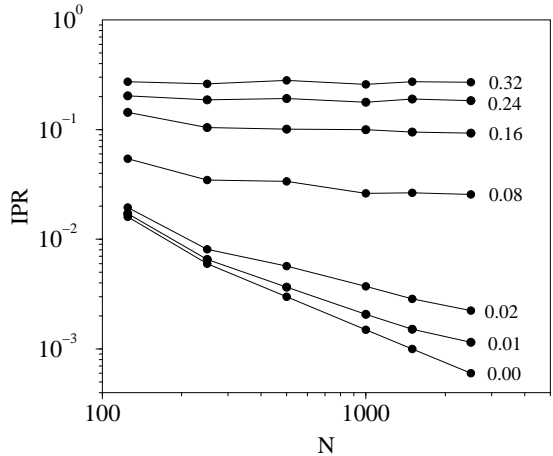,width=8cm}}
\caption{Scaling of the inverse participation ratio with system size
for the eigenstates at the center of the band. Labels indicate the
degree of disorder.}
\label{fig5}
\end{figure}

As mentioned above, the scaling of the IPR with the lattice size may be
useful to discern the nature of the eigenstates.  The IPR at the center
of the band for different values of the degree of disorder is plotted in
Fig.~\ref{fig5} as a function of the lattice size.  The IPR for periodic
lattices scales very accurately as $1/N$, hence indicating that their
eigenstates spread uniformly over the whole lattice.  As soon as some
amount of disorder is introduced in the system, the IPR follows a power
law for small $N$ but tends to a constant value for large $N$, as
plotted in Fig.~\ref{fig5}.  The critical size for which deviation from
power fit occurs decreases on increasing the degree the disorder.  The
constant value of the IPR for large $N$ increases with the degree of
disorder, indicating that the eigenstates at the center of the band
actually become more localized.

Summarizing these observations for the NN approximation, we are led to two
main conclusions.  First, the zero-energy peak of the DOS really exists.
Moreover, as its width shows no dependence on the degree of disorder (at
least, when the latter ranges over the interval used in our simulations),
we are inclined to identify this peak with a $\delta$-like singularity
rather than to the famous Dyson singularity $\sim 1/|E|\ln^3|E|$.  This
singularity was found firstly for a special distribution of the NN hopping
integral in the form of a generalized Poisson function.\cite{Dyson53}
Second, the corresponding eigenstates show no tendency to delocalization
with rising the lattice size contrary to the opposite statement done in
Ref.~\onlinecite{Theodorou76}.  Moreover, they are not more delocalized
than those of energies close to zero.

\subsection{All interactions} \label{All}

Effects of inclusion of all dipolar interactions in Eq.~(\ref{eigen}) on
the DOS and IPR has been already discussed in
Ref.~\onlinecite{Fidder91}.  Nevertheless, it has been done only for a
fixed values of the chain length $(N=250)$ and the degree of disorder
$(\sigma=0.08)$.  Below we present our DOS and IPR numerical data
obtained by varying both $N$ and $\sigma$.

\begin{figure}
\centerline{\epsfig{file=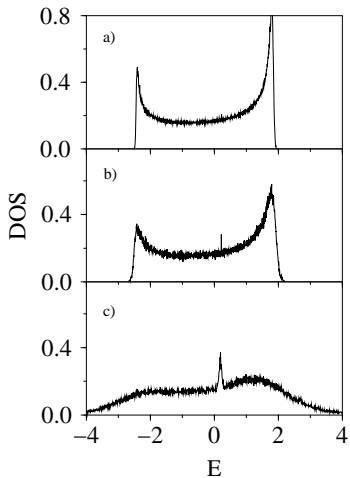,width=8cm}}
\caption{Density of states in the frame of LR coupling when the
lattice size is $N=2500$ and the degree of disorder is a)
$\sigma_{NN}=0.02$, b) $0.04$, and c) $0.16$.}
\label{fig6}
\end{figure}

\begin{figure}
\centerline{\epsfig{file=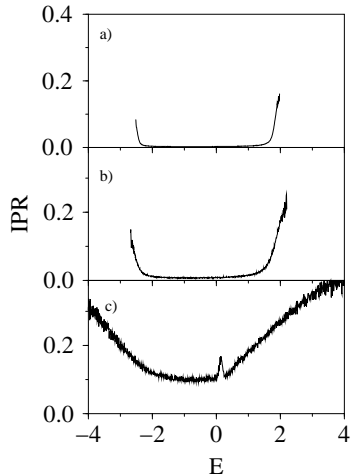,width=8cm}}
\caption{Inverse participation ratio for the same cases shown in
Fig.~\protect{\ref{fig6}}.}
\label{fig7}
\end{figure}

Figures~\ref{fig6} and \ref{fig7} show the results of numerical
calculations of both the DOS and the IPR for different values of the
degree of disorder and $N=2500$.  Here, one can observe the usual
changes of both magnitudes as compared to those in the NN approximation:
asymmetry and shift of the excitonic band edges, both in a good
agreement with the analytical results presented in Sec.~\ref{A}.

In addition, some new features appear, namely the peak in the DOS has a
finite width and is shifted from zero-energy to a somewhat higher value
$E_{\mathrm{peak}}\approx 0.21U$ for low degree of disorder, in full
correspondence with the results of numerical simulation done in
Ref.~\onlinecite{Fidder91}.  Higher values of the degree of disorder
lead to a smaller energy shift.  Further, a peak in the IPR appears at
the same energy where the DOS peak, with a finite width as well.  The
last observation means that the states forming the DOS peak become more
localized as compared to those with close energies, in contrast to the
case of the NN interaction.  This is also confirmed by the scaling of
the IPR for $E_{\mathrm{peak}}$ with the system size (not shown here):
In all cases we observe higher values of the IPR in comparison with
those shown in Fig.~\ref{fig5}.  Besides that, the trend is similar,
that is, the IPR scales as $1/N$ only for perfect lattices, whereas it
tends to a constant value for non-zero degree of disorder.

\subsection{Discussion} \label{Dissc}

Now let us discuss the origin of the features of the DOS and of the IPR
found in numerical simulations.

\subsubsection{NN interaction} \label{NN_int}

Obviously, the zero-energy peak in the DOS might appear when the states
at the center of the exciton band become localized, i.e., their
localization lengths are reduced to values less than the lattice size.
This occurs when the reduced degree of disorder due to motion narrowing,
$\sigma_{nd}(0,0)= 6\sigma_{NN}U/(N+1)^{1/2}$, exceeds the energy
spacing at the center of the band, $\Delta E=2\pi U/(N+1)$.
Equalizing these two magnitudes one obtains an estimation for a {\em
threshold\/} of mean fluctuations of the NN distances to observe the
peak, $\sigma_{NN}^{th} \approx U/(N+1)^{1/2}$.  Thus $\sigma_{NN}^{th}
\approx 0.02$ for $N=2500$, which is in a good agreement with the
numerical data of Fig.~\ref{fig1}.

With regard to the fact why this peak appears, we can suggest two
explanations which seem to be suitable for the model under consideration.
First, as the distribution of disorder we used has long tails then, owing to
possible large fluctuations of the NN distances, strongly interacting {\em
dimers\/} can be created whose level splittings noticeably exceed the
typical magnitude of the intersite interaction $U$.  Consequently, the
whole chain is broken into several independent segments in the sense that
two adjacent {\em dimers\/} produce a potential well for the exciton,
localizing it into the segment bounded by them.  As the zero eigenenergy is
always present in a segment with odd number of sites, one can expect a peak
in the DOS at this energy (similar explanation of the zero-energy peak of
the DOS was suggested in Ref.~\onlinecite{Lifshits82}).  The peak amplitude
increases with disorder simply because of the rising of the number of
segments as the degree of disorder grows.  Appearance of such strongly
interacting dimers is clearly seen from that the IPR approaches 0.5 at the
DOS tails (see Figs.~\ref{fig3} and~\ref{fig4}).

Recently, it was demonstrated that the Dyson singularity of the DOS
appeared even for a box-like distribution of disorder.\cite{Mertsching92}
Then, the explanation above fails due to the absence of large fluctuations
of the NN-randomness at a low magnitude of the degree of disorder.  In such
a case, another cause for the occurrence of the zero-energy peak in the DOS
can be proposed.  As we have already noted in subsection~\ref{B}, the first
order correction to the central energy is equal to zero for chains with odd
number of sites and has a minimum of fluctuation in the case of even number
of sites.  The zero-energy peak indicates that the central band
eigenenergies are more stable to perturbations than the remaining ones.
This certainly will result in a peak of the DOS after averaging over
realizations of disorder.  It is remarkable that simulations done for a
special type of disorder ---which has no effect on a certain excitonic
level in the sense that the first order correction to the energy
vanishes--- show an analogous peak in the DOS at this energy.\cite{Kozlov}
Thus, this {\em empirical\/} rule can serve for inspecting the appearance
of peaks in the DOS for the tight-binding Hamiltonian.  As the last
treatment does not use any specific peculiarities of the NN-randomness
distribution, it seems to be suitable for any other distribution.  We
suppose that, for the model considered in this paper, both mechanisms
discussed above contribute to the formation of the zero-energy peak in the
DOS.

Concluding this subsection, note that the degree of localization of the
central states obtained from the numerical simulation is in a good
agreement with the theoretical estimates based on a self-consistent rule
proposed in Refs.~\onlinecite{Malyshev95} and~\onlinecite{Malyshev91}
(see paragraph 1 of the present subsection).

\subsubsection{All interactions} \label{All_int}

As it was stated in Ref.~\onlinecite{Fidder91}, the energy shift of the
DOS peak, $E_{\mathrm{peak}}\approx 0.21 U$, agrees very well with the
energy of the central band state in the absence of disorder,
Eq.~(\ref{Epeak}).  We are also inclined to relate this peculiarity to a
state of analogous origin, i.e., similar to $\sin(\pi n/2)$.  This can
be demonstrated at least in the perturbative limit.  Moreover,
exploiting this analogy further, we should assume that the character
(having no amplitude on the half of sites) of the mentioned eigenstate
has not to be changed dramatically (at least, in average), when passing
from the NN model to the exact one, as it is the case for the problem
without disorder.\cite{Malyshev95}

The $\delta$-singularity of the DOS becomes broader with including all
dipolar couplings.  At least two effects can contribute to this
broadening.  As it was supposed above, the DOS peak results from the
isolated segments of odd number of sites, which, in turn, originate from
large fluctuations of the NN distances.  At moderate magnitudes of
disorder we are mainly dealing with, it is unlikely the simultaneous
strong reduction of the distance between a nearest-neighbor pair and the
distances with other neighbors.  Thus, for the very beginning, one can
consider adjacent segments as independent of each other.  Then, the
eigenenergy of the local (belonging to a certain segment) central band
state will fluctuate, owing to fluctuations of the segment lengths [see
Eq.~(\ref{Epeak})], and thus will produce inhomogeneous broadening of
the DOS peak.  The second probable origin of this effect is the coupling
of different isolated segments due to the interaction with far
neighbors.

In Ref.~\onlinecite{Fidder91}, the appearance of the IPR peak was
explained by an exceptional property of this characteristic with regard
to the central band state, $k = (N+1)/2$, characterized by the wave
function $[2/(N+1)]^{1/2}\sin(\pi n/2)$.  Even in the absence of
disorder, the IPR defined by Eq.~(\ref{IPR}) shows stronger localization
of this state $[{\cal L} = 2/(N+1)]$ as compared to localization of the
remaining states $[{\cal L} = 3/2(N+1)]$.\cite{Fidder91} The authors of
Ref.~\onlinecite{Fidder91} asserted that the IPR peak in the presence of
disorder reflected a remnant of this special state in those forming the
peak.  At the moment, we do not see any other explanation of the origin
of this anomaly.  If it is so, the similar feature might be manifested
in the NN problem, too.  Nevertheless, as follows from our simulations
done for the NN problem, the IPR displays a downfall rather than a peak.
One of the reason for such difference may be the fact that the
zero-energy state is not exponentially localized in the NN problem (see
Sec.~\ref{?}).  It results in a larger extension of this state as
compared to the others.  In principle, such a large extension can
compensate the IPR anomaly coming from the special character of the
zero-energy state (having no amplitude at all on the half of sites)
giving rise to the same value of the IPR at $E=0$ and at surrounding
energies.

\section{Summary} \label{Sum}

The numerical study of the problem of the zero-energy peak of the DOS
for a one-dimensional Frenkel chain with only off-diagonal randomness
shows that the peak is really presented.  In the NN approximation, it is
located at the center of the excitonic band and tends to convert to a
$\delta$-singularity as the size of the chain increases.  The states
belonging the peak are localized and do not display any tendency to
delocalization with the chain size.  Moreover, the degree of
localization (IPR) does not differ very much from that of the
surrounding states.  The inclusion of couplings due to far neighbors
shifts the peak to a slightly higher energy ($\approx 0.21 U$), while
the IPR, in contrast to the NN problem, shows a peak at the same energy.

\acknowledgments

Work at Russia was supported by the Russian Foundation for Basic
Research (Project 95-03-09221).  Work at Madrid was supported by CICYT
under Project MAT95-0325.  V.\ A.\ M.\ thanks Universidad de Salamanca,
where this study was started, for hospitality.

\end{document}